# Villager's dilemma

**Beihang He**
**Advertisement 0702,Department of Arts and Communication, Zhejiang University City College**
**No.51 Huzhou Street Hangzhou Zhejiang, 310015**
**Tel: 13906539819**
**E-mail:** hebeihang@gmail.com

**Abstract:**With deeper study of the Game Theory, some conditions of Prisoner's Dilemma is no longer suitable of games in real life. So we try to develop a new model−Villager's Dilemma which has more realistic conditions to stimulate the process of game. It is emphasize that Prisoner's Dilemma is an exception which is lack of universality and the importance of rules in the game. And it puts forward that to let the rule maker take part in the game and specifies game players can stop the game as they like.

This essay describes the basic model, the villager's dilemma (VD) and put some extended use of it, and points out the importance of rules and the effect it has on the result of the game. It briefly describes the disadvantage of Prisoner's Dilemma and advantage Villager's Dilemma has. It summarizes the premise and scope of application of Villager's Dilemma, and provides theory foundation for making rules for game and forecast of the future of the game.

## *1. Basic Model*

In the basic model, the villager's dilemma (VD) is presented as follows:
Three villagers who have the same physical strength and a robber who has two and a half times physical strength as three villagers are living in the same village. In other words, three villagers has to act together to defeat the robber who is stronger than anyone of them.
Generally, each villager has 5 bags of grain produced, and they have to suffer from depredation of the robber every year.
The robber follows the rule that he will take away three bags of grain from every villager per year in ordinary conditions. However, the villager who defies or resists the robber will be despoiled of all the production, and the villager who betrays the other two villagers who prefer to betray and acts as an accomplice will be rewarded of two bags of grain without any loss. If there are two betrayers, then each of them will be given one bag of grain.
It is known that three villagers can communicate freely without being held back by the robber.
If villagers prefer to defy or resist the robber just once, in other words, they choose to defeat the robber together, and then they will play the game repeatedly. And if the game is unavoidable, the game will become a game played between villagers and the village chief, which is an extended model detailed in Chapter Two.

| | Villager A & Villagers B Defy | Villager A & Villagers B obey | Villager A Defies, Villagers B Betrays | Villager A Betrays, Villagers B Defies |
|---|---|---|---|---|
| Defy | 5 bags of grain | 0 bags of grain | 0 bags of grain | 0 bags of grain |
| Obey | 2 bags of grain | 2 bags of grain | 2 bags of grain | 2 bags of grain |
| Betray | 7 bags of grain | | 7 bags of grain | 7 bags of grain |

Table1. Strategy for Villager C in Villager's Dilemma. (Assuming that each villager is rational)
In Villager's Dilemma, three villagers represent game players of three parties. And the robber stands for the rule maker. Villagers are free to choose their own strategies. Meanwhile, the robber is able to make rules as he likes. This model is similar to the prisoner's dilemma at some point, while it is more approach to the real situation. We will discuss this in Chapter Four.
Compared with other dilemma theory, we also assume that there are no reputaton effects from the villager's decision. And the only concern of each player (villager) is maximizing his/her own payoff, without concern for the other player's payoff. The unique equilibrium for this game is a Pareto-suboptimal solution-that is, rational choice leads the two players to both play defect even each player's individual reward would be greater if they both played cooperatively.

Which strategy does a villager should to choose if they want to minimize his/her own loss and maximize payoff?
In this game, given that three villagers communicate with each other to defy before robber's coming, one of them will consider, if I choose the strategy to defy, then the optimal strategy for the other two is to betray. Then I will lose all my production because of my defiance. However, if I choose to help the robber, no matter what the other villager does, it will be better than defy. So that, betray and acts as an accomplice is the optimal strategy.
It is clear that three villagers are confront with the same reality, and each of them will arrive at the conclusion that to betray and acts as an accomplice is the optimal strategy based on rational pondering. That is to say the Subgame perfect Nash equilibrium is all the three villagers obey and become accomplice of the robber. Under this circumstance, each villager loses three bags of grain, while robber gets the maximum payoff.
Nash equilibrium of the game above is evidently not the Pareto Optimum. Concerning of the payoff of the three villagers as whole, if all of them choose to defy and defeat the robber, they will remain all their production—five bags of grain per person, which is a better payoff for them, other than betray in which there is just two bags of grain remained for each person. But based on the hypothesis, the three rational individuals concern of each own payoff without other's. Equilibrium of this game is three villagers choose the same strategy to obey, then payoff for each one is better than defy, but payoff for the whole is low. It's what is called 'dilemma'.

## *2. Extended model.*

### 2.1 Result of Game Given that Abolish the Reward rule.

Let's omit the rule that ' the villager who betrays the other two villagers who prefer to betray and acts as an accomplice will be rewarded of two bags of grain without any loss'. Then we will get the result as following:

| | Villager A & Villagers B Defy | Villager A Obeys, Villagers B defies | Villager A & Villagers B obey |
|---|---|---|---|
| Defy | 5 bags of grain | 0 bags of grain | 0 bags of grain |
| Obey | 2 bags of grain | 2 bags of grain | 2 bags of grain |
| Betray | 2 bags of grain | 2 bags of grain | 2 bags of grain |

Table2. Strategy for Villager C in Villager's Dilemma. (Assuming that each villager is rational)

At this time, the Nash equilibrium alters: three villagers will choose the strategy of defying after communicating. And the Nash Equilibrium and the Pareto Optimum arrive at the same point.

## 2.2 Result of Game Given that Abolish the Punishment rule.

We have found that the result of the game alters if we omit the reward rule. Then if we omit the punishment rule, what will happen?

| | Villager A & Villagers B Defy | Villager A & Villagers B obey | Villager A Defies, Villagers B Betrays | Villager A Betrays, Villagers B obeys |
|---|---|---|---|---|
| Defy | 5 bags of grain | 2 bags of grain | 2 bags of grain | 2 bags of grain |
| Obey | 2 bags of grain | 2 bags of grain | 2 bags of grain | 2 bags of grain |
| Betray | 7 bags of grain | | 6 bags of grain | 7 bags of grain |

Table2. Strategy for Villager C in Villager's Dilemma. (Assuming that each villager is rational)

The most remarkable phenomenon of this strategy is that the optimal strategy is to defy without communication among them, by reason that the act of betrayal and the act of defiance must occur together. Payoff will be reduced if one chooses to obey, while it will be increased if chooses to defy.

At the same time, we found that after communication, the result is not the same as before at all. Thus, three villagers will regard betray as the optimal strategy for individual! It is sure that every rational individual will not communicate with each other in advance to make other known his/her option, since the betray strategy is on the basis of knowing others will defy. However, we know that defy is the optimal strategy without communication.

Under the circumstance it is not possible to attain a table result. Therefore, we are sure to found defying is the optimal strategy after communication in the repeated game in long-run.

Why result of the extended model does not correspond with Prisoner's Dilemma. The reason is that we omit the rule of reward and punishment, which actually alters the final result.

The reward and punishment system is used to encourage game players to choose the strategy which the rule maker wishes them to choose other than strategies which is prohibited. At this point, there're obvious differences occur between the game players and the rule maker:

1. If there is reward, player will choose to 'betray and act as accomplice'.
2. If there is neither reward nor punishment, player will choose to 'obey'.
3. If there is punishment, player will choose to 'betray and act as accomplice'.

We have found that, if only the robber make rules which are easy enough to be distinguished, Nash Equilibrium of the Prisoner's Dilemma will be achieved. It is evidentially that Villager's Dilemma is a special case of Prisoner's Dilemma. Furthermore, a number of conditions, such as prisoners cannot communicate with each other, rule maker will not be affected by the game players, the rule will stay the same are more and more inadaptable to the society.

## 2.3 Repercussion the alteration of rule will have on Villager's Dilemma

If we define the rule maker as boss other than robber, and use the profit made by employees, payment for boss and welfare for employees to replace the grain, then the game will have a fundamental change. Under this circumstance, rational game player and rule maker will concern of how to realize the profit maximization for collectivity instead of individuals.

At this moment, rule maker and game players have changed their moods. To be more precise, it means that boss has to provide high payment and proper welfare for employees to prevent their job-hoping because it is necessary to employees. And the boss will set rules as following:

If you do a bad job, then you will get less payment.

If you do an average job, then you will get the basic payment.

If you do a good job, then you will get extra payment.

Above is a simple model of modern corporation system. From this model we can see that you will never get rid of bosses' restraint. And you have no choice but to bargain with your boss or move to a better job. In addition, this game is doomed to be played repeatedly in your career.

In this game, everyone should not just concern of himself or herself. We all know that if the boss pays a low payment, then he will suffer from employees' job-hopping. And if an employee asks for a very high payment, he will be fired as well, or else, the corporation will be shut up for the too heavy income burden. To sum up, result of seeking personal interests is the loss of all the people's good.

It is predicable that an equilibrium of the game between boss and employees will be arrived with the more games played, and a mutual benefit system will be established.

In order to make the game continued, boss ought to hew to the rule:

Rules must be worked out reasonably, and there are significant differences between reward and punishment.

It is notable that it is required to set rules which have marked differences. First, it is necessary to estimate and analyze game players in all aspects in order to prevent errors in setting differences. For example, there is no difference to villagers that the robber robs three bags or all production of villagers if villagers need three gags of grain to survive. And at this time, the optimal strategy can not be anything but defiance. It is the same in corporate management, we have to set wage standard based on local general level of market price.

## *3. Examples and disadvantages of Prisoner’s Dilemma.*

*Before talking about the usage of Villager’s Dilemma in real-life, let’s review the classic form of Prisoner’s Dilemma in Game Theory.*

*The following part is quoted from Wikipedia:*

*The Prisoner's Dilemma constitutes a problem in game theory. It was originally framed by Merrill Flood and Melvin Dresher working at RAND in 1950. Albert W. Tucker formalized the game with prison sentence payoffs and gave it the "Prisoner's Dilemma" name (Poundstone, 1992).*

*In its classical form, the prisoner's dilemma (PD) is presented as follows:*

*Two suspects are arrested by the police. The police have insufficient evidence for a conviction, and, having separated both prisoners, visit each of them to offer the same deal. If one testifies (defects) for the prosecution against the other and the other remains silent, the betrayer goes free and the silent accomplice receives the full 10-year sentence. If both remain silent, both prisoners are sentenced to only six months in jail for a minor charge. If each betrays the other, each receives a five-year sentence. Each prisoner must choose to betray the other or to remain silent. Each one is assured that the other would not know about the betrayal before the end of the investigation. How should the prisoners act?*

*If we assume that each player prefers shorter sentences to longer ones, and that each gets no utility out of lowering the other player's sentence, and that there are no reputation effects from a player's decision, then the prisoner's dilemma forms a non-zero-sum game in which two players may each cooperate with or defect from (i.e., betray) the other player. In this game, as in all game theory, the only concern of each individual player (prisoner) is maximizing his/her own payoff, without any concern for the other player's payoff. The unique equilibrium for this game is a Pareto-suboptimal solution—that is, rational choice leads the two players to both play defects even though each player's individual reward would be greater if they both played cooperatively.*

*In the classic form of this game, cooperating is strictly dominated by defecting, so that the only possible equilibrium for the game is for all players to defect. No matter what the other player does, one player will always gain a greater payoff by playing defect. Since in any situation playing defect is more beneficial than cooperating, all rational players will play defect, all things being equal.*

*In the iterated prisoner's dilemma the game is played repeatedly. Thus each player has an opportunity to punish the other player for previous non-cooperative play. If the number of steps is known by both players in advance, economic theory says that the two players should defect again and again; no matter how many times the game is played. Only when the players play an indefinite or random number of times can cooperation be an economic equilibrium. In this case, the incentive to defect can be overcome by the threat of punishment. When the game is infinitely repeated, cooperation may be a subgame perfect Nash equilibrium although both players’ defecting always remains equilibrium and there are many other equilibrium outcomes.*

*In casual usage, the label "prisoner's dilemma" may be applied to situations not strictly matching the formal criteria of the classic or iterative games; for instance, those in which two entities could gain important benefits from cooperating or suffer from the failure to do so, but find it merely difficult or expensive, not necessarily impossible, to coordinate their activities to achieve cooperation.*
*The classical prisoner's dilemma can be summarized thus:*

| | *Prisoner B Stays Silent* | *Prisoner B Betrays* |
|---|---|---|
| *Prisoner A Stays Silent* | *Each serves 6 months* | *Prisoner A: 10 years*<br>*Prisoner B: goes free* |
| *Prisoner A Betrays* | *Prisoner A: goes free*<br>*Prisoner B: 10 years* | *Each serves 5 years* |

*In this game, regardless of what the opponent chooses, each player always receives a higher payoff (lesser sentence) by betraying; that is to say that betraying is the strictly dominant strategy. For instance, Prisoner A can accurately say, "No matter what Prisoner B does, I personally am better off betraying than staying silent. Therefore, for my own sake, I should betray." However, if the other player acts similarly, then they both betray and both get a lower payoff than they would get by staying silent. Rational self-interested decisions result in each prisoner's being worse off than if each chose to lessen the sentence of the accomplice at the cost of staying a little longer in jail him. Hence it is a seeming dilemma. In game theory, this demonstrates very elegantly that in a non-zero sum game a Nash Equilibrium need not be a Pareto optimum.*
*Prisoner's Dilemma has been quoted for many times, 'in a non-zero sum game a Nash Equilibrium need not be a Pareto optimum' is regarded as philosophy by people. But we also find some problem which cannot be explained by Prisoner's Dilemma with deeper exploration.**[1]***
Let's analyze two examples below:

## Case 1

*William Poundstone, in a book about the Prisoner's Dilemma (see References below), describes a situation in New Zealand where newspaper boxes are left unlocked. It is possible for people to take a paper without paying (defecting) but very few do, feeling that if they do not pay then neither will others, destroying the system. Because there is no mechanism for personal choice to influence others' decisions, this type of thinking relies on correlations between behaviors, not on causation. Because of this property, those who do not understand superrationality often mistake it for magical thinking. Without superrationality, not only petty theft, but voluntary voting requires widespread magical thinking, since a non-voter is a free rider on a democratic system.* ***[2]***

## Case 2

*When cigarette advertising was legal in the United States, competing cigarette manufacturers had to decide how much money to spend on advertising. The*

*effectiveness of Firm A's advertising was partially determined by the advertising conducted by Firm B. Likewise, the profit derived from advertising for Firm B is affected by the advertising conducted by Firm A. If both Firm A and Firm B chose to advertise during a given period the advertising cancels out, receipts remain constant, and expenses increase due to the cost of advertising. Both firms would benefit from a reduction in advertising. However, should Firm B choose not to advertise, Firm A could benefit greatly by advertising. Nevertheless, the optimal amount of advertising by one firm depends on how much advertising the other undertakes. As the best strategy is dependent on what the other firm chooses there is no dominant strategy and this is not a prisoner's dilemma but rather is an example of a stag hunt. The outcome is similar, though, in that both firms would be better off were they to advertise less than in the equilibrium. Sometimes cooperative behaviors do emerge in business situations. For instance, cigarette manufacturers endorsed the creation of laws banning cigarette advertising, understanding that this would reduce costs and increase profits across the industry. This analysis is likely to be pertinent in many other business situations involving advertising.* ***[3]***

## Analysis of cases

For case one, let's talk about it together with some real fact:
Modern media corporation always use advertising strategy as profit mode rather than the sale of newspapers. And we find more and more newspapers begin to decrease the price of a single newspaper, even deliver newspapers freely in order to make a high price on the basis of a high broad reach.
In New Zealand, the credit history is very demanding. Once there is a stain in the credit history, it is difficult for one to do anything smoothly.
From A we find that payoff is quite low and even can be ignored if we choose betray. However, from B it is clear that punishment is very painful for anyone. In Villager's Dilemma, it is can be described as changing the conditions in the game, making the mode as following:

| | **Prisoner B Stays Silent** | **Prisoner B Betrays** |
|---|---|---|
| **Prisoner A Stays Silent** | Each serves 10 months | Prisoner A: 2 years<br>Prisoner B: 5 years |
| **Prisoner A Betrays** | Prisoner A: 5 years<br>Prisoner B: 2 years | Each serves 10 years |

Now both parties choose to collaborate, and the game reaches Nash Equilibrium. This result has disproved the deduction that 'in a non-zero sum game a Nash Equilibrium need not be a Pareto optimum'.How did the wrong deduction be produced? The result is that Prisoner's exclude rules in the game. Actually, rules will have significant effect on the result of the game.
Case two has improved that rules will have significant effect on the result of the game with real-world example. And we have to remember that the rule has to be efficient rather then something useless.

Let's discuss two questions:
Replace the newspapers in Case One with a Lamborghini of the latest version.
Replace the laws in Case Two with Intra-industry agreements.
Results:
Cannot be reach collaboration because payoff of betray is extremely large.
Cannot be reach collaboration because punishment is not available when Intra-industry agreements exist.
These results can be found in real life. And it pointed out that Prisoner's Dilemma has an obviously problem:
Prisoner's Dilemma has set fixed rules with distinguishing reward and punishment, but it is exception in real life. Prisoner's Dilemma overlook the importance of rules and effect rules have on the result of games, only considered strategies chosen by game players.

Why do some corporation choose betray which is at high risk and may be punished by Opponent rivals? We know that demand in a given period is constant. For example, the more the people to divide a pie, the less amount everyone will get. On the contrary, the less the people to divide a pie, the more amount everyone will get. That is to say, with increase of the game players, payoff will be also rise for everyone, and finally make game players tend to choose betraying. It is more appropriate in the real word, but payoff is fixed for every game player in Prisoner's Dilemma disagree with the situation in real life.

## *4. Advantage of Villager's Dilemma.*

The most significant advantage that Villager's Dilemma has over Prisoner's Dilemma is that it is more approach to the virtual reality. Specifically, we have found seven advantages:
There're distinguishable differences between strategies. Betray is segmented into two groups—Betray and act as accomplice which can be rewarded and obey which cannot get any rewards.
The less the game layers, the more payoff will betrayer to attain.
Game players are allowed to communicate with each other.
Rule maker also takes part in allocation of benefit.
Any participant of the game have right to stop the game.
The model with three game players is easier to be used in the real-life with more than two parties in the game.
Setting rules has vital effect on result of the game.

## *5. Practical significance of the model of villager's dilemma.*

Villager's Dilemma is of interest to the Sociology, Politics, Management and Economics.

## 5.1 Result of the Game will be affected by the Rules.

In Prisoner's Dilemma, there's doubt of Adam Smith's Invisible Hand Theory. Now, it's high time to complete the conclusion. According to the demonstration above, we add one point to Adam Smith's Theory.

**In the market economic with has distinguishable reward and punishment ,that is to say, one will be awarded because of behaviors benefit the party, otherwise he or she will be punished because of behaviors harm the party, everyone will finally make the party's benefit to the maximum based on the consideration of his or her own.**

Powerful rule maker make rules to affect game players' decision, he tries to control the result—the invisible hand, that is the market economic by making rules—the visible hand, such as laws, regulations and policies.

It should be pointed out that, **rules must be effective, and consideration of reward and punishment must be in related to all game players.**

Rules which are ineffective include some examples below:

1. Rules made by electronic appliance holding in private.
2. Credit History without morality.
3. Laws of the country in war.

We have mentioned that rules must be effective, and consideration of reward and punishment must be in related to all game players. It means that if a person needs at least three bags of grain to survive, then the rule is not well present reward and punishment.

The classical example is Case Two which tells that alter rules can affect result of game in an extremely extent. Besides this example, we can also find other case in the real life.

## 5.2 Games which lack of Distinguishing Rules cannot be Played Smoothly.

*A record from Wanxiang Cave, China, characterizes Asian Monsoon (AM) history over the past 1810 years. The summer monsoon correlates with solar variability, Northern Hemisphere and Chinese temperature, Alpine glacial retreat, and Chinese cultural changes. It was generally strong during Europe's Medieval Warm Period and weak during Europe's Little Ice Age, as well as during the final decades of the Tang, Yuan, and Ming Dynasties, all times that were characterized by popular unrest. It was strong during the first several decades of the Northern Song Dynasty, a period of increased rice cultivation and dramatic population increase. The sign of the correlation between the AM and temperature switches around 1960, suggesting that anthropogenic forcing superseded natural forcing as the major driver of AM changes in the late 20th century.[4]*

From this article we find that all dynasties are characterized by that natural calamity happened and lead to a general region of famine when one is week and dying. However, is that a dynasty will overthrow if only there's famine? In the light of history, there were frequent natural disasters before West Han Dynasty. But because emperors cared of people and the country, it weren't cause the unrest of society, instead, it raised emperor's reputation.

Thus, what's the common ground of dynasties' decaying? The answer is political corruption.

Let's have a look at Villager's Dilemma's use in this case:

Because of natural disaster's, there was a bad harvest, every villager produced three bags of grain. However, officials are greedy, and pocketed food for victims, at the same time, put harsh duties on people, and took five bags of grain away from every people.

In fact, villagers and robbers were not allowed to communicate with each other under the feudal system. Finally, villagers defied and ended up the game, so that the dynasty decayed.

The Chinese Nation was considering the question: Why did people always suffer from bitter life either the country was strong or poor? Whereas America which was on the on the other side of the ocean established democratic system to get rid of the odd circle. In other words, American democratic system provided the channel for villagers and robber to communicate with each other, and robber was selected by villagers. Furthermore, villagers decay via voting rather than military force, which we called an peaceful and efficient way to resolve a contradiction.

## *6. Conclusion*

Villager's Dilemma is aimed at analyze how rules made by managers will affect the result of the game of managed game players. It emphasizes the importance of rules' making. This mode provides explain and solutions of the current situation.